\documentclass[superscriptaddress,prl,twocolumn,tightenlines,10pt,a4paper]{revtex4}%
\usepackage{bbm}
\usepackage{amsfonts}
\usepackage{amssymb}
\usepackage{graphicx}
\usepackage{subfigure}
\usepackage{savesym}
\usepackage{amsmath}
\usepackage{txfonts}
\usepackage{multirow}
\usepackage{epstopdf}
\usepackage{color}

\newcommand{\ket}[1]{\vert#1\rangle}
\newcommand{\bra}[1]{\langle#1\vert}

\begin{document}

\title{Generation of quantum discord between ionic qubits via noisy processes}\vspace{-6pt}

\author{B. P. Lanyon}
\author{P. Jurcevic}
\author{C. Hempel}

\affiliation{Institut f\"ur Quantenoptik und Quanteninformation,\\
\"Osterreichische Akademie der Wissenschaften, Technikerstr. 21A, 6020 Innsbruck,
Austria}

\affiliation{
Institut f\"ur Experimentalphysik, Universit\"at Innsbruck,
Technikerstr. 25, 6020 Innsbruck, Austria}

\author{M. Gessner}

\affiliation{
Department of Physics, University of California, Berkeley, California 94720, USA}
\affiliation{
Physikalisches Institut, Universit\"at Freiburg, Hermann-Herder-Strasse 3, D-79104 Freiburg, Germany}

\author{V. Vedral}
\affiliation{
Centre for Quantum Technologies, National University of Singapore, 3 Science Drive 2, 117543, Singapore}
\affiliation{
Clarendon Laboratory, Department of Physics, University of Oxford, Parks Road, Oxford OX1 3PU, U.K.}
\affiliation{Department of Physics, National University of Singapore, 2 Science Drive 3, 117542, Singapore
}

\author{R. Blatt}
\author{C. F. Roos}
\affiliation{Institut f\"ur Quantenoptik und Quanteninformation,\\
\"Osterreichische Akademie der Wissenschaften, Technikerstr. 21A, 6020 Innsbruck,
Austria}

\affiliation{
Institut f\"ur Experimentalphysik, Universit\"at Innsbruck,
Technikerstr. 25, 6020 Innsbruck, Austria}

\date{\today}

\begin{abstract} 

Quantum systems in mixed states can be unentangled and yet still correlated in a way that is not possible for classical systems.  These correlations can be quantified by the quantum discord and might provide a resource for certain mixed-state quantum information processing tasks. Here we report on the generation of discordant states of two trapped atomic ions via Markovian decoherence processes. While entanglement is strictly non-increasing under such operations, discord can be generated in various forms. Firstly we show that, starting from two classically correlated qubits, it is possible to generate discord by applying decoherence to just one of them. Secondly, even when starting with completely uncorrelated systems, we show that discord can be generated via classically correlated decoherence processes. Finally, the Werner states are created. The generated states can be used as a resource state for quantum information transmission and could be readily extended to more ions.  

\end{abstract}

\maketitle

Over the last decade improvements in experimental control over quantum systems have enabled scientists to generate non-classical states of light and matter. Entanglement between quantum systems \cite{RevModPhys.81.865} has been at the forefront of this research largely due to the key role it plays in Bell inequalities \cite{bell} and as a resource for certain information processing tasks \cite{MikeIke}. However, entanglement is not the only kind of non-classical correlation that can exist between systems. 
Discord was introduced as a measure of these other correlations in bipartite systems, which becomes distinguishable from entanglement for mixed states \cite{0305-4470-34-35-315, PhysRevLett.88.017901, datta:050502, review}. There are mixed states with discord yet no entanglement and these are the focus of the present work.  

There is growing evidence that, like entanglement, discord can be viewed as a resource for information processing tasks \cite{Dakic:2012fk, teleportation, Gu:2012uq, PhysRevLett.81.5672,datta:042316, datta:050502}. 
Recently, quantum states of light have been generated which belong to the regime where there is discord but no entanglement \cite{PhysRevLett.101.200501, PhysRevLett.109.030402, Dakic:2012fk,Gu:2012uq}. Following extension of the discord concept to continuous variable systems \cite{PhysRevLett.105.030501, PhysRevLett.105.020503}, experiments have also investigated Gaussian discord dynamics under various decoherence channels \cite{PhysRevLett.109.030402}. Here we are concerned with discrete variable systems (qubits).

This paper is organised as follows: After briefly reviewing the definition of discord we demonstrate how operations on one qubit can generate discord between two qubits, via a simple example. Secondly, we review the correlation rank as a way to assess the nature of correlations in quantum states and then show how discordant states with any rank can be generated via noisy processes. Finally, we present the first generation of the Werner states between atomic qubits.

\begin{figure}[t]
\vspace{-5mm}
\includegraphics[width=1 \columnwidth]{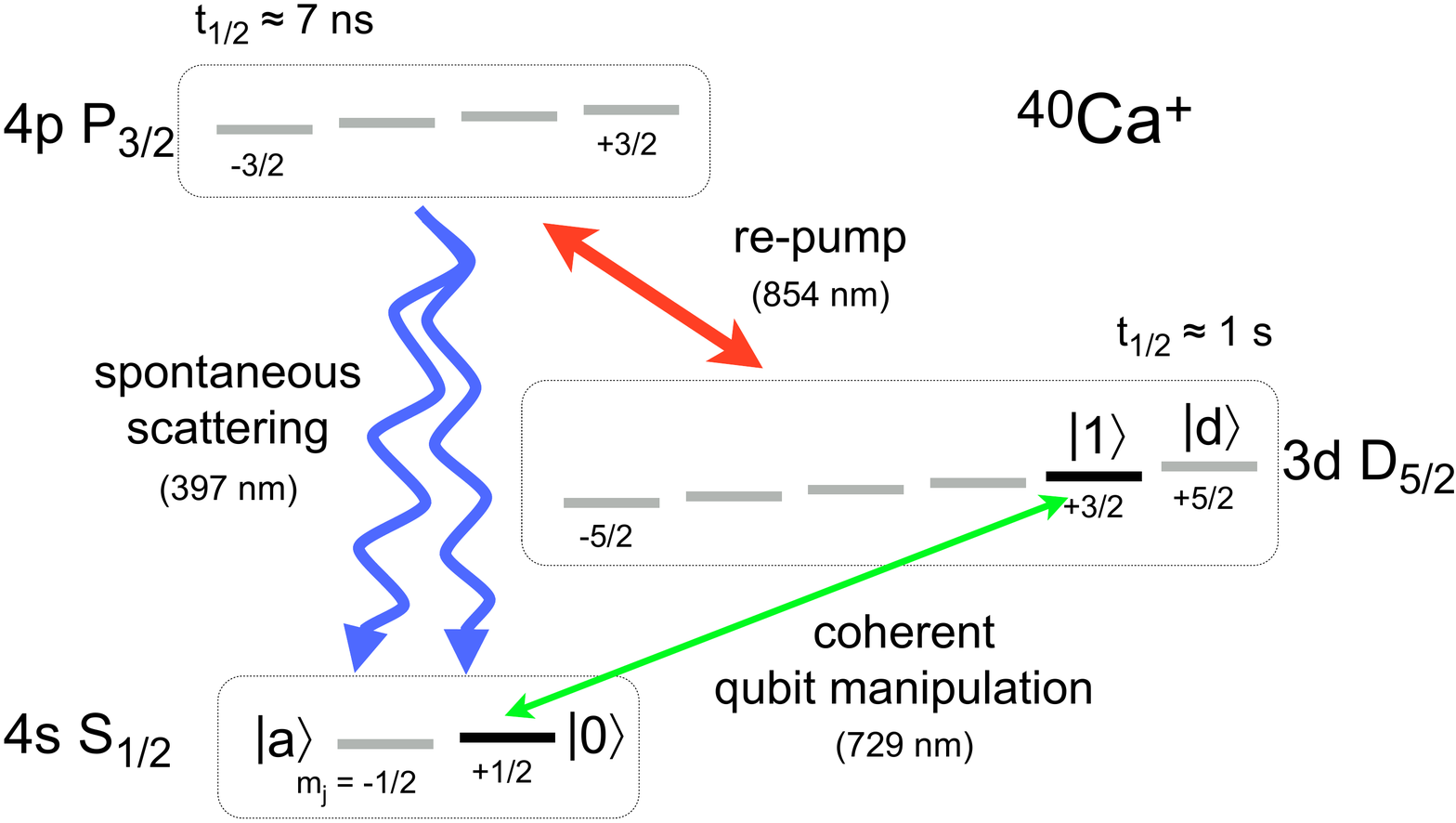} 
\vspace{-8mm}
\caption{\label{traps}
Potential energy level scheme of the $^{40}$Ca$^{+}$ ion showing the relevant Zeeman sublevels.
A qubit is encoded in the states $S_{1/2}(m_j{=}+1/2){=}\ket{0}$ and $D_{5/2}(m_j{=}+3/2){=}\ket{1}$, where $m_j$ is the magnetic quantum number. 
ancillary levels $\ket{a}$ and $\ket{d}$ are temporarily employed for state preparation, as described in the text. 
The qubit is at optical frequencies (729 nm), the re-pump is infrared (854 nm) and the decay channel is UV (397 nm).  Adjacent Zeeman states in the $S$ and $D$ manifolds are shifted by about $10$~MHz. 
}
\label{levelscheme}
\vspace{-4mm}
\end{figure}

Two quantum systems A and B have discord, when considering measurements on system A, if and only if their state cannot be written in the form $\rho_{AB}{=}\sum_{i}p_i\ket{\psi_i^A}\bra{\psi_i^A}\otimes \rho_{i}^{B}$, where $\bra{\psi_i}\psi_j\rangle{=}\delta_{ij}$, $\rho_{i}^{B}$ are density matrices of qubit B and $p_i$ are probabilities. For a state $\rho_{AB}$ a von Neumann measurement \cite{MikeIke} of A with eigenvectors $\Pi_i{=}\ket{\psi_i}\bra{\psi_i}$ will leave the total state unchanged, i.e. $\sum_i \Pi_i\rho_{AB}\Pi_i^{\dagger}{=}\rho_{AB}$. The discord $D$ of a bipartite system is quantified as the difference between two definitions of the mutual information, $I$ and $J$, i.e. $D{=}I{-}J$. The first definition ($I$) captures the total correlations of the density matrix $\rho_{AB}$ by the difference in entropy of systems when taken individually and when taken together $I(\rho_{AB}){=}S(\rho_A)+S(\rho_B)-S(\rho_{AB})$, where $S$ is the von Neumann entropy \cite{MikeIke}  and $\rho_{A}(\rho_{B})$ is the reduced density matrix of system $A$ ($B$). The second definition ($J$) captures the classical correlations and can be interpreted as information gain about one subsystem as a result of a measurement on the other. $J(\rho_{AB}){=}S(\rho_B){-}S(\rho_{B|A})$, where $S(\rho_{B|A})$ is the entropy of B after a measurement of A (with unknown result) and $J$ is maximised over all von Neumann measurements of A, see \cite{review}. Discord can be asymmetric with respect to exchange of the two systems since it quantifies the extent to which measurements on one system affects the total system for an independent observer. For the discord that considers measurements on system A (B) we will use the label $D_A$ ($D_B$). 

Consider the separable two-qubit mixed state
\begin{equation}
\rho_1=\frac{1}{2}\left(\ket{+}_A\bra{+}\otimes \ket{+}_B\bra{+}~ ~+~~ \ket{-}_A\bra{-}\otimes \ket{-}_B\bra{-}\right),
\end{equation}
where $\ket{\pm}{=}(\ket{0}{\pm}\ket{1})/\sqrt{2}$. Although manifestly correlated, state $\rho_1$
 is fully classical since it is diagonal in the local orthogonal $\ket{\pm}\otimes\ket{\pm}$ basis. The discord is correspondingly zero in either direction: a von Neumann measurement $\{\Pi_{\pm}=\ket{\pm}\bra{\pm}\}$ of either qubit leaves the density matrix unchanged. 
Surprisingly, discord can be generated by applying an amplitude damping process that acts \emph{only} on one of the qubits in state $\rho_1$ \cite{PhysRevA.85.010102,PhysRevLett.107.170502}. 

Amplitude damping of a single qubit can be described by the quantum map $\epsilon^\prime_{ad}(\rho){=}E_{0}\rho E_0^\dagger{+}E_1\rho E_1^\dagger$ with Kraus operators $E_0{=}\ket{0}\bra{0}{+}\sqrt{1-p}\ket{1}\bra{1}$ and $E_1{=}\sqrt{p}\ket{0}\bra{1}$. For $p{=}0~(p{=}1)$ the qubit undergoes  zero (complete) amplitude damping. 
Consider now the case where after preparation of $\rho_1$ qubit B undergoes this process, i.e. it interacts with a dissipative Markovian bath that causes the excited state $\ket{1}$ to decay to the ground state $\ket{0}$. 
Throughout the damping process the state is of the form $\epsilon_{ad}(\rho_1){=}\frac{1}{2}(\ket{+}_A\bra{+}\otimes \tau_{+B}+\ket{-}_A\bra{-}\otimes  \tau_{-B})$, where  $\tau_{\pm B}{=}\epsilon_{ad}^\prime(\ket{\pm}\bra{\pm})$ are generic density matrices representing the state of qubit $B$. 
The effect of the damping process is to reduce the distinguishability of $\tau_{\pm B}$, which are initially orthogonal. 
Subsequently, qubit A becomes correlated with non-orthogonal states of qubit $B$. For $0{<}p{<}1$ there is no von Neumann measurement of $B$ after which the original state, $\epsilon_{ad}(\rho_1)$,  is recovered and therefore the discord $D_{B}{>}0$. 

Experiments are carried out using a system of two $^{40}$Ca$^+$ ions in a linear Paul trap. A qubit is encoded in an $S_{1/2}$ ground and a $D_{5/2}$ metastable state, which are coupled by a narrow quadrupole transition (see Fig.~\ref{levelscheme}). Correlations are established between the state of the valence electrons in two ions. To generate the complex correlated mixed states additional electronic levels are employed to both temporarily store information and remove entropy from the system. 

\begin{figure}
\vspace{-3mm}
\includegraphics[width=1 \columnwidth]{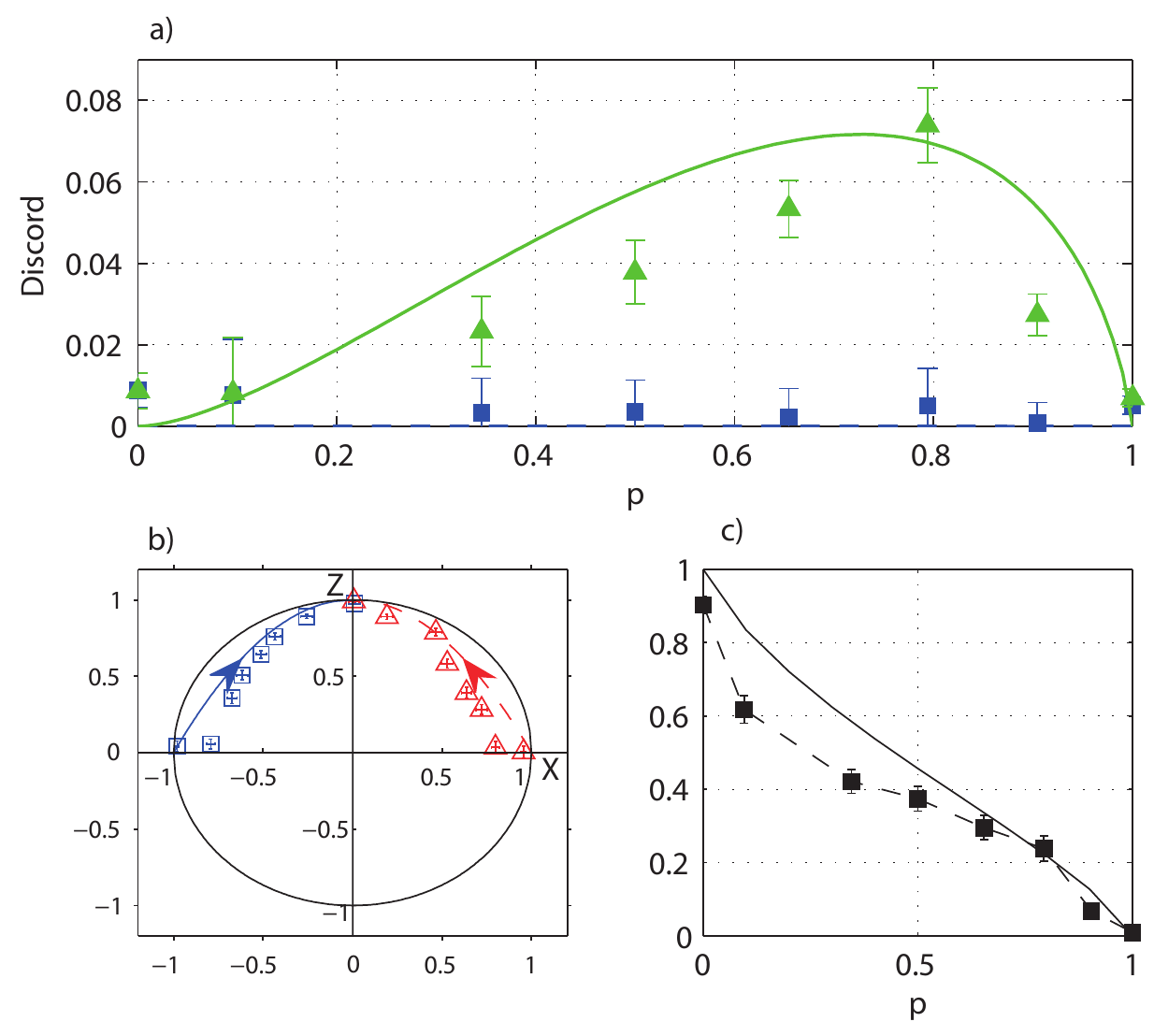}
\vspace{-7mm}
\caption{\label{figure2}
Amplitude damping results $\epsilon_{ad}(\rho_1)$ derived from experimentally reconstructed density matrices.  
(a) Discord $D_{A}$ (ideal dashed blue line, data squares) and $D_{B}$ (ideal green line, data triangles) as a function of amplitude damping probability $p$ (see text). 
(b) Ideal trajectories $\tau_{+ B}{=}\epsilon_{ad}^\prime(\ket{+}\bra{+})$ (blue solid line) and $\tau_{- B}{=}\epsilon_{ad}^\prime(\ket{-}\bra{-})$ (red dashed line) on the X-Z plane of the Bloch sphere. Open shapes with error bars show experimental results derived from the reduced state of qubit $B$ in Eq.~(1) after projecting qubit $A$, in the reconstructed density matrix, into either $\ket{+}$ (triangles) or $\ket{-}$ (squares) i.e. $\mathrm{Tr}_{A}(\ket{\pm}\bra{\pm}\epsilon_{ad}(\rho_1))$. 
(c) Total correlations correlations, $I(\rho_{AB}){=}S(\rho_A)+S(\rho_B)-S(\rho_{AB})$ (see text). Ideal (solid black line) and experimental (solid squares with error bars).}
\label{figure2}
\vspace{-3mm}
\end{figure}

Firstly, the initial classically-correlated state $\rho_1$ (Eq.~1) is prepared. Using a standard optical pumping technique, both qubits are initialised into $\ket{0}$ (Fig.~\ref{levelscheme}). The two qubit unitary operation $MS(\theta{=}\pi/4){=}\exp(-i\theta \sigma_x^1\sigma_x^2)$ is applied, where $\sigma_x$ is a Pauli spin-1/2 operator,  generating a maximally entangled state $(\ket{00}{-}i\ket{11})/\sqrt{2}$ \cite{PhysRevLett.82.1971}. This operation is realised using a narrow-band bichromatic laser pulse with frequency components close to the first red and blue axial centre-of-mass vibrational sidebands of the two-ion crystal, as described in detail in \cite{Kirchmair:2009}. A 20~kHz symmetric detuning from the sidebands is employed, entangling the qubits in 50~$\mu s$. Next, a series of resonant 729~nm pulses moves this superposition into the state $(\ket{dd}{-}i\ket{aa})/\sqrt{2}$. An 854~nm laser pulse then couples the $\ket{d}$ state to the short-lived $P_{3/2}$ state ($t_{1/2}{\approx}7~ns$) which, due to selection rules, can only spontaneously decay into $\ket{0}$ --- thereby preparing the mixed state $(\ket{aa}\bra{aa}{+}\ket{00}\bra{00})/2$. Finally, a 729~nm pulse on the $\ket{a}{\leftrightarrow}\ket{1}$ transition, followed by another on the qubit transition, prepares the target state Eq.~(1). In summary, the classical correlations in $\rho_1$ are established by first generating quantum correlations, then removing all the phase information via an irreversible process on both ionic-qubits. 

A local amplitude damping channel, with probability $p$, is realised by moving the corresponding fraction of qubit B's population from $\ket{1}$ into the ancillary state $\ket{a}$, by a tightly-focused 729~nm laser beam. Next, a circularly polarised 397~nm laser pulse incoherently transfers the $\ket{a}$ population into $\ket{0}$. 
Ramsey experiments show that coherent superpositions in the qubit transition are largely unaffected by this 397~nm `optical pumping' pulse. The total experimental time is approximately 500~$\mu$s,  of which 460~$\mu$s are required to generate the initial state Eq. (1).

Full state tomography of the generated two-qubit state is carried out, for a range of amplitude damping probabilities~$p$. The maximum likelihood reconstruction method \cite{PhysRevA.64.052312} is employed, providing the most likely physical state to have produced the measurement data. Errors are determined via a Monte Carlo simulation of projection noise (due to a finite number of measurements) centred around the experimentally recorded data. 1000 measurements were taken in each basis to convincingly distinguish between zero and non-zero discord states. The set of zero discord two-qubit states is of zero measure in the total set \cite{PhysRevA.81.052318} and therefore any white noise (e.g. projection noise) is likely to result in the reconstruction of a discordant state. For detailed numerical simulations of this effect see Supplemental Material. All quantities are derived from the reconstructed density matrix, including discord via the usual numerical optimisation over projective measurements \cite{PhysRevLett.88.017901, PhysRevLett.101.200501}. 

Results from the amplitude damping experiment are summarised in Fig.~\ref{figure2}. Figure~\ref{figure2}a shows the discord observed in each direction. The results clearly demonstrate the generation of statistically significant amounts of discord $D_B$, while $D_A$ remains zero to within error. In all cases the reconstructed states contain no entanglement, quantified by the tangle \cite{PhysRevA.61.052306}, to within one standard deviation (not shown). 
Figure~\ref{figure2}b presents the dynamics of the states of qubit B, $\tau_{\pm B}$, on a cross section of the Bloch sphere. The distinguishability is reduced by the damping process and consequently, for all $0{<}p{<}1$, qubit $A$ becomes correlated with non-orthogonal states of qubit B. 

Our results do not imply that the total correlations between two systems can be increased via operations on only one system. 
Fig.~\ref{figure2}c shows that the total correlations, captured by the mutual information $I$,  continually decrease during the damping channel. The correct interpretation is that the process converts some of the preexisting classical correlations, quantified by $J$, into quantum correlations. Indeed, if there are no classical correlations present in the initial state, then no discord can be generated by operating on only one system \cite{PhysRevA.85.052122}. Furthermore, only a very restricted class of discordant states can be created by such operations: a set of measure zero in the total set \cite{PhysRevA.85.052122}. That there are fundamentally different kinds of discordant two-qubit states, in terms of how they can be generated, raises the question of whether there is another way of quantifying the correlations in these systems.

 \begin{figure}
\includegraphics[width=1 \columnwidth]{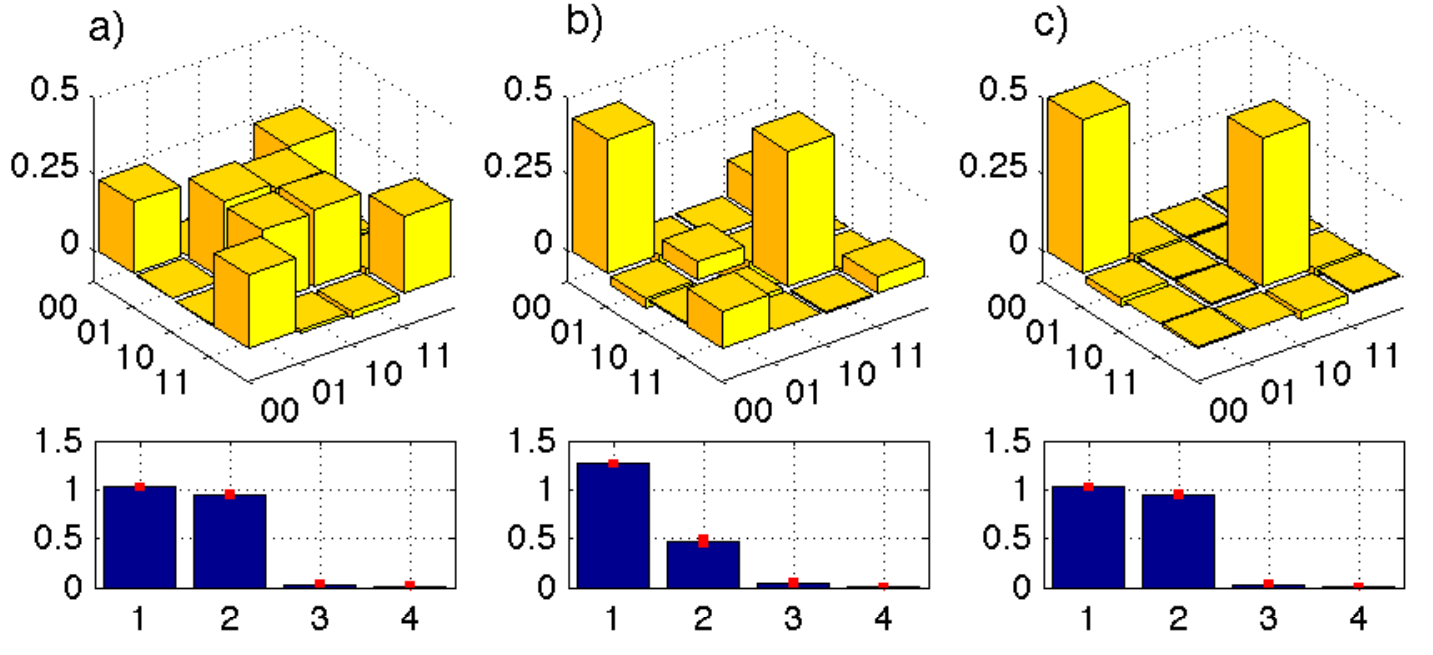}
\vspace{-5mm}
\caption{\label{figure3}
Amplitude damping results. Real values of experimentally reconstructed density matrices for damping probabilities ($p$) (a) 0.00 (b) 0.79 and (c) 1.00. Imaginary components are $\leq$ 0.03. Below, bar charts show the singular values of the corresponding correlation matrix. The number of non-zero values gives the rank. The fidelities with target states are a) 0.984$\pm{0.004}$ b) 0.992$\pm{0.006}$ and c) 0.991$\pm{0.002}$. The tangle (a measure of two qubit entanglement \cite{PhysRevA.61.052306}) of each state is zero to within error (not shown). 
}
\vspace{-4mm}
\end{figure}

An alternative view on quantum correlations is presented in Ref.~\cite{PhysRevA.85.052122}, where in addition to discord, the rank $R$ of the correlation matrix is considered. Although operations on one system can convert classical to quantum correlations, they \emph{cannot} increase the correlations in terms of $R$, which is therefore considered as an additional quantity of interest. $R$ is calculated in the following way. First, the correlation matrix (M) is constructed by writing $\rho$ in a basis of local Hermitian operators (e.g. the Pauli spin operators). $R$ is then the number of non-zero singular values of M. Hence, $R$ gives the minimal number of bipartite product operators needed to represent a given quantum state. Originally, $R$ was introduced as a witness for discord \cite{PhysRevLett.105.190502}. A state with $R{=}1$ is completely uncorrelated. If $R> d_A$, where $d_A$ is the dimension of the smallest system, then the discord is non-zero. For a system of two qubits $d_A{=}d_B{=}2$. The lowest rank of a two-qubit system containing discord is 2, but in general an $R{=}2$ state may or may not contain discord. The maximum is $R{=}4$, representing a highly correlated state. In Ref.~\cite{teleportation} it is shown that $R$ determines the extent to which a state can act as a resource in quantum state transmission. 

Fig.~\ref{figure3} presents a selection of reconstructed states, and their singular values, from the amplitude damping experiment. The results are largely consistent with an invariant value $R{=}2$. Small non-zero singular values consistent with a higher rank are analysed in the Supplemental Material. Numerical simulations show that these small but somewhat statistically significant values (larger than zero by less than 2-3 standard deviations) are consistent with the effects of measurement `projection' noise, even with the large numbers of measurements employed. The difficulties associated with verifying that singular values are strictly zero clearly makes the witness criteria for discord very challenging experimentally.   

How then can these strongly correlated high rank discordant states be generated? We now show that classical noise processes which operate on both qubits are sufficient to generate states of all ranks. Consider the following: the qubits interact with an environment which causes both qubits to suffer identical single-qubit rotations, around some axis, by an amount that is not known and fluctuates from experiment to experiment. Complete dephasing under this classically correlated noise can be modelled by a quantum map $\epsilon_{cd}^{\vec{n}}{=}\frac{1}{2\pi}\int_0^{2\pi} K_{\vec{n}}(\theta)\rho K_{\vec{n}}^{\dagger}(\theta)$ with operators $K_{\vec{n}}(\theta){=}R_{\vec{n}}(\theta){\otimes}R_{\vec{n}}(\theta)$ and $R_{\vec{n}}(\theta){=}e^{-i\theta\vec{n}\cdot\vec{\sigma}/2}$ is a single qubit rotation around a normalized axis vector $\vec{n}$. The integral over the angle $\theta$ generates a dephasing effect between eigenstates of the rotation operator. 

This type of noise occurs in any experimental situation with fluctuating external fields that couple equally to the qubits. In the case of trapped-ion qubits, correlated changes in the coupling strength between lasers and ions can occur due to intensity or beam-pointing fluctuations, for example. Other prime candidates are the fluctuation of ambient electric or magnetic fields which remain constant over the closely-spaced ion string.

We investigated the effects of correlated dephasing due to ambient magnetic (B) field noise. Although these effects are always present, typical experiments require orders of magnitude less time than that required for complete dephasing. Here we intentionally expose various quantum states for long times. Dephasing via B-field fluctuations can be understood as follows.  The transition frequency of each qubit is determined by the local B-field, which is largely identical for both ions due to their small separation (${\approx} 5~\mu$m). 
B-field noise causes correlated qubit rotations in the Z basis, i.e. $\vec{n}\cdot\vec{\sigma}{=}\sigma_z$, and $\theta$ is proportional to the B field. Over the many repeated experimental runs required to estimate expectation values for quantum state reconstruction, the effect of these fluctuations is to smear out the reconstructed phases between eigenstates of the $\sigma_z$ operators (logical states). (As previously shown \cite{PhysRevLett.106.130506}, when applied to certain entangled GHZ states this noise process can be extremely detrimental, resulting in a fully classical $R{=}2$ state. Here we show the converse---the process can also generate quantum correlations). 

\begin{figure}
\vspace{-5mm}
\includegraphics[width=1 \columnwidth]{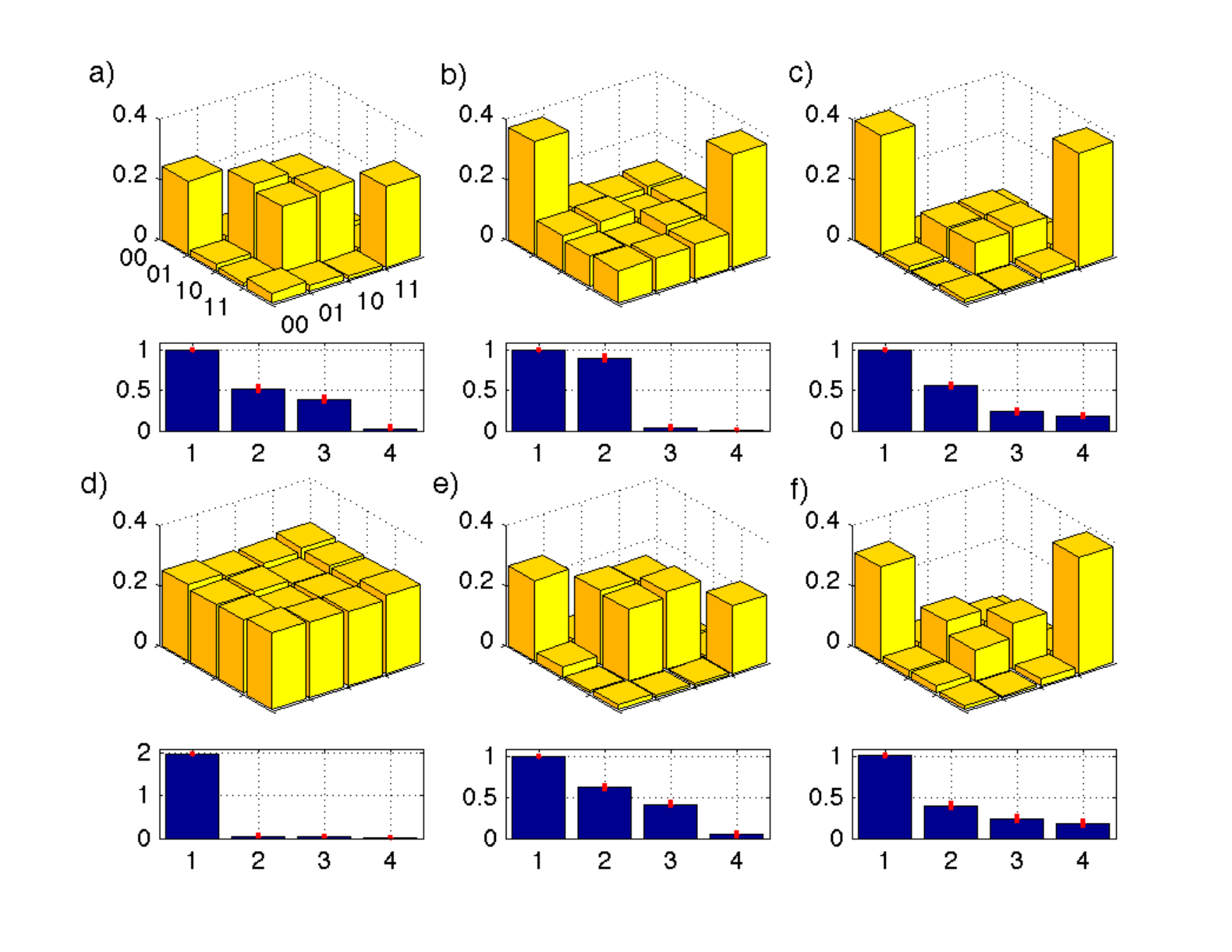}
\vspace{-5mm}
\caption{\label{figure4}
Correlated dephasing results.  Real values of experimentally reconstructed density matrices, and corresponding singular values of the correlation matrix (underneath), for target states:   
a) $\epsilon_{cd}^{\vec{n}}(\rho_1)$; 
b) $\rho_{2}$ (see text); 
c) $\epsilon_{cd}^{\vec{n}}(\rho_{2})$; 
d) $\ket{\psi}{=}\ket{+,+}$, where $\ket{+}{=}(\ket{0}{+}\ket{1})/\sqrt{2}$; 
e) $\epsilon_{cd}^{\vec{n}}(\ket{\psi}\bra{\psi})$;
f) $\epsilon_{cd}^{\vec{n}}[K_{\vec{m}}(\pi/2)\,\epsilon_{cd}^{\vec{n}}(\ket{\psi}\bra{\psi})\,K_{\vec{m}}(\pi/2)^{\dagger}]$, for $\vec{m}\cdot\vec{\sigma}{=}\sigma_y$ and $\vec{n}\cdot\vec{\sigma}{=}\sigma_z$.
The discord $D_{B}$ is a) 0.19$\pm{0.03}$ b) 0.01$\pm{0.01}$ c) 0.188$\pm{0.002}$ d) 0.011$\pm{0.005}$ e) 0.22$\pm{0.03}$  f) 0.12$\pm{0.03}$.
The tangles are all zero to within error and the discord $D_A$ is equal to $D_{B}$ to within error in each case.
Imaginary components of density matrices are all $\leq$ 0.05.
}
\vspace{-3mm}
\end{figure}

The dephasing channel is applied to the classically correlated $R{=}2$ state $\rho_1$, by introducing a delay between initialisation of the qubits into this state and performing tomographic measurements. A delay of 10~ms is found to be sufficient to achieve almost complete dephasing, which is over two orders of magnitude longer than that required to perform coherent laser-driven qubit operations. Experimentally reconstructed density matrices and singular values of $\rho_1$ and $\epsilon_{cd}^{\vec{n}}(\rho_1)$ are shown in Fig 3.a) and Fig 4.a), respectively for $\vec{n}\cdot\vec{\sigma}=\sigma_z$. One observes the almost complete loss of the $\ket{00}\bra{11}$ coherence element of the density matrix \cite{ban}, while the $\ket{01}\bra{10}$ coherence is largely unaffected since it is in a decoherence-free subspace \cite{chwalla}. The measured discord increases from $0.009{\pm}{0.004}$ to $0.19{\pm}{0.03}$ and the singular values are consistent with an increase from an $R{=}2$ to an $R{=}3$ state. 

In the Supplemental Material we present new theoretical results on the conditions under which correlated noise can change the correlation rank R of a quantum state and by how much. We show that the final rank can be obtained with the aid of a simple geometrical picture, expressing the relationship between the rotation axis $\vec{n}$ of the correlated dephasing and two normalized vectors $\vec{v},\vec{w}\in\mathbb{R}^3$ which provide all the necessary information about $R{=}2$ states. Specifically, $R{=}2$ states where the reduced state of each qubit is completely mixed (e.g. state $\rho_1$) can be written as $\rho=\frac{1}{4}\left(\mathbb{I}\otimes\mathbb{I}+d\vec{v}\cdot\vec{\sigma}\otimes\vec{w}\cdot\vec{\sigma}\right)$. The final rank depends on the overlap between $\vec{n}$ and the vectors $\vec{v}$ and $\vec{w}$, respectively. In $\rho_1$ the qubits are correlated in the $X$ direction, i.e., $\vec{v}=\vec{w}=\vec{e}_x$ and the dephasing rotations are conducted around the orthogonal $Z$ direction, $\vec{n}=\vec{e}_z$. In this case an $R{=}3$ state is generated since $\vec{n}\cdot\vec{v}=\vec{n}\cdot\vec{w}=0$. However if $\vec{n}$ is neither equal nor orthogonal to $\vec{v}$ and $\vec{w}$ then an $R{=}4$ state can be generated. We demonstrate this by preparing the new  initial state $\rho_{2}{=}K_{\hat{n}}(\pi/8)\rho_1K_{\hat{n}}(\pi/8)^{\dagger}$ for $\vec{n}\cdot\vec{\sigma}{=}\sigma_y$, by applying a 729~nm laser pulse $K_{\hat{n}}(\pi/8)$ to $\rho_1$.  Figure~\ref{figure4}(b-c) present the results for this initial and final state: the significantly increased 3\textsuperscript{rd} and 4\textsuperscript{th} singular values are consistent with the conversion from an $R{=}2$ to an $R{=}4$ state.

In all cases considered so far classical correlations have been present in the initial state. However, we also find that completely uncorrelated $R{=}1$ states ($\rho{=}\rho_A{\otimes}\rho_B$) can be converted to high rank states (see Supplemental Material). The conditions for this can be described in terms of the reduced Bloch vectors of the two qubits $\vec{r}^A$ and $\vec{r}^B$. If either $\vec{n}{=}\vec{r}^A$ or $\vec{n}{=}\vec{r}^B$ then $R$ is unchanged by correlated noise. In any other case an $R{=}3$ state is generated. Figure~\ref{figure4}(d-e) presents results that demonstrate the conversion from an $R{=}1$ to $R{=}3$ state. Fig.~\ref{figure4}(f) also shows that even conversion from $R{=}1$ to $R{=}4$ is possible, by effectively employing a sequence of dephasing operations. Specifically, as detailed in Fig.~\ref{figure4}, we allow the the $R{=}1$ state to fully decohere via correlated B-field noise into an $R{=}3$ state, apply a unitary rotation and then allow the state to fully decohere in the same way again. 

Finally we report on the generation of Werner states \cite{PhysRevA.40.4277} $\rho_W(p){=}p\ket{\phi^+}\bra{\phi^+}{+}(1{-}p)\mathbb{I}/4$, which are a mixture of a maximally entangled  Bell state $\phi^+{=}(\ket{00}{+}\ket{11})/\sqrt{2}$ and the maximally mixed two-qubit state $\mathbb{I}/4$. These states have rank $R{=}4$ for all $p{>}0$ and have been studied in a number of different contexts \cite{PhysRevA.64.030302}. A range of different Werner states have previously been made in light fields \cite{PRA65012301, PhysRevLett.92.177901}. 
 For $p{\leq}1/3$ the Werner states are separable, for $p{>}1/3$ they are entangled. In contrast, the discord in $\rho_W(p)$, which is symmetric under exchange of the particles ($D_A{=}D_B$), runs smoothly from zero at $p{=}0$ to a maximum at $p{=}1$.

\begin{figure}[h]
\vspace{0mm}
\includegraphics[width=1 \columnwidth]{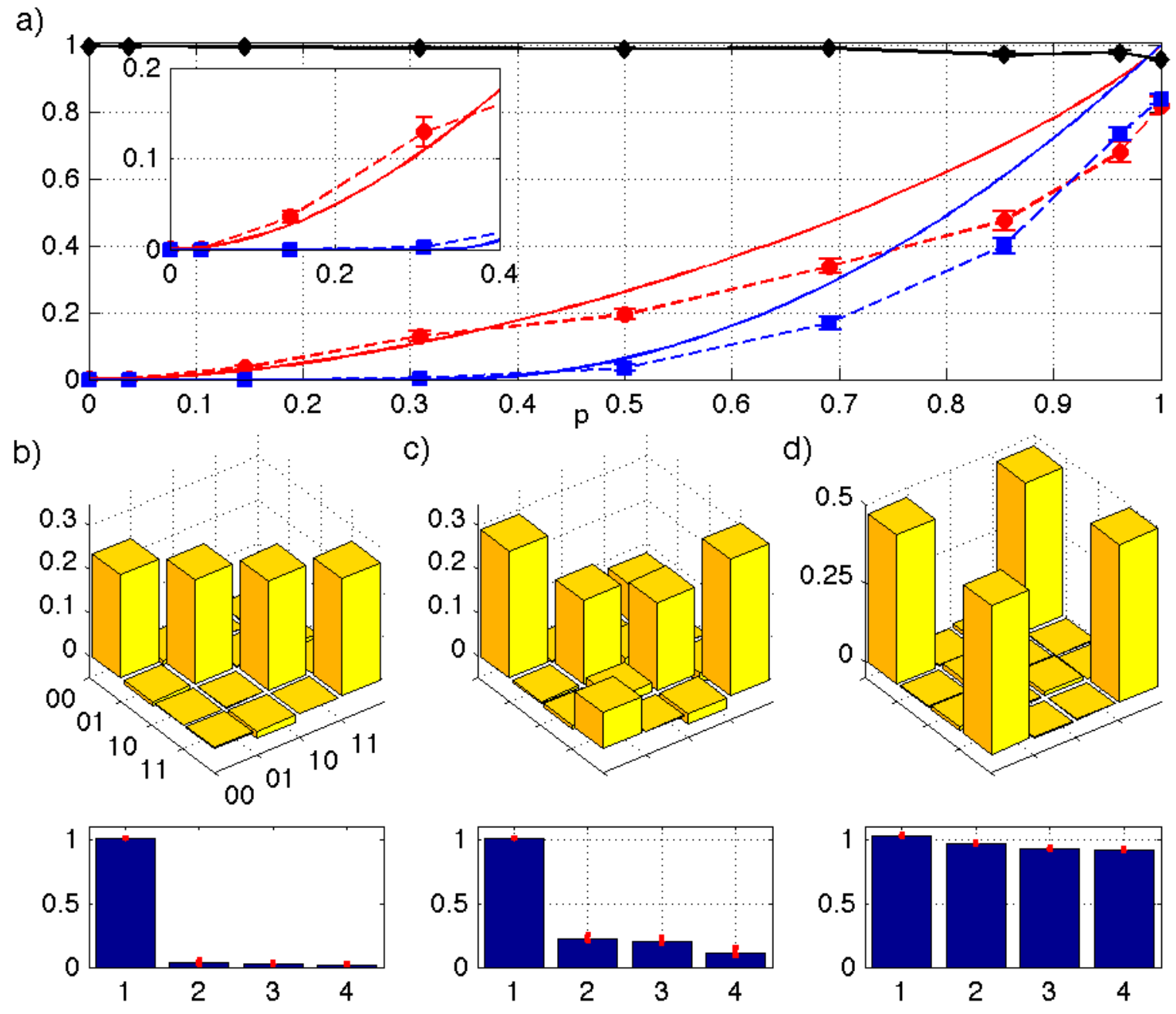}
\vspace{-6mm}
\caption{\label{traps}
Werner state results. 
a) Measured discord $D_{A}$ (red circles), tangle (blue squares), fidelity with the ideal Werner state (black diamonds). Solid lines show ideal values. Inset shows smooth increase in discord, while the tangle (a measure of two qubit entanglement \cite{PhysRevA.61.052306}) begins abruptly at $p{=}1/3$. The discord $D_{B}$ is equal to $D_{A}$ to within error in each case. 
b) Real values of experimentally reconstructed density matrices measured for b) $p{=}0$, c) $p{=}0.31$ and  d) $p{=}1.00$.  Imaginary values are all less than $0.03$. 
Below each matrix the singular values of the corresponding correlation matrix are shown.
}
\label{figure5}
\vspace{-6mm}
\end{figure}

Werner states were made by applying the operation $MS2{=}\exp(-i~\pi/4~\sigma_{\j}\sigma_{\j})$, where $\sigma_{\j}{=}(\sigma_{x}{+}\sigma_{y})/\sqrt{2}$, to the classically correlated input state $\rho_{W}^{in}{=}p\ket{00}\bra{00}{+}(1{-}p)\mathbb{I}$. The state $\rho_{W}^{in}$  is prepared in the following way: 1)~the qubits are entangled via $MS2(\cos^{-1}[\sqrt{p}])\ket{00}{=}\sqrt{p}\ket{00}{+}\sqrt{1-p}\ket{11}$: 2) an 854~nm laser pulse with suitable polarisation causes spontaneous decay of the $\ket{11}$ population, via the $P_{3/2}$ state, into the states $\ket{a}$ and $\ket{0}$ with equal probability---thereby generating a state equivalent to $\rho_{W}^{in}$ but with the mixed state across the $\ket{0}{\leftrightarrow}\ket{a}$ transition: 3) A 729~nm pulse on the $\ket{a}{\leftrightarrow}\ket{1}$ transition prepares $\rho_{W}^{in}$. Experimental results are presented in Fig.~\ref{figure5}. The observed states achieve a high fidelity with ideal case. The discord is found to increase smoothly from zero, while the entanglement begins abruptly close to the ideal value of  $p{=}1/3$. 

We have shown that in stark contrast to entanglement, discord can be generated between two systems via operations on just one system. 
Not all discordant states can be made this way, and the correlation rank provides a way to distinguish between different kinds of discordant states. 
Finally, we have shown that noise processes generated by classically fluctuating fields are sufficient to generate discordant states with any rank, even starting with completely uncorrelated states. 

We gratefully acknowledge support by the European Commission via the integrated project AQUTE and a Marie Curie Fellowship (PIIF-GA-2010-275477) supporting BPL. VV acknowledges support from the National Research Foundation and the Ministry of Education, Singapore as well as the James Martin School (UK), Leverhulme Trust (UK), Engineering and Physical Sciences Council (UK) and the Templeton Foundation (USA). MG thanks the German National Academic Foundation for support.


\newpage
\clearpage

\section{SUPPLEMENTARY MATERIAL}

\subsection{Effects of quantum projection noise on tomographic state reconstruction}

In this section we present results from numerical simulations of the effects of measurement noise on experimentally reconstructed density matrices. Specifically, we ask the question ``if the ideal state is generated in the lab what are the effects of taking only a finite set of measurements (i.e. measurement or projection noise) on the reconstructed density matrices". The results show that even for very large numbers of measurements (a thousand repeated measurements per observable) the states reconstructed via maximum likelihood tomography have statistically significant differences to the ideal states. 

State tomography requires experimentally estimating the probabilities for finding the quantum system(s) to be in various states. This must be done by making a finite number of repeated measurements. If $n$ trials are made and $n_i$ instances of the outcome $i$ occur then the assigned probability is $p_i{=}n_i/n$. For the expected uncertainty in this value we use the Binomial proportion confidence interval for one standard deviation $\sqrt{p_i(1-p_i)/n}$.

We simulate the effects of a finite number of measurements $n$ following a Monte Carlo approach:
Taking the probability distribution of a perfect state we make a number of copies (70) and add in each case noise sampled randomly from binomial distributions defined by the ideal probabilities and $n$, the number of samples. We then reconstruct the density matrix of each noisy data set via maximum likelihood tomography. The mean and standard deviation of a range of properties, like the discord and entanglement for example, can then be calculated and compared to the ideal values.

As an example we take the ideal fully classically correlated and separable state given in equation~(1) in the main text:

\begin{equation}
\rho_1=\frac{1}{2}\Big(\ket{+}_A\bra{+}\otimes \ket{+}_B\bra{+} \;\, + \;\, \ket{-}_A\bra{-}\otimes \ket{-}_B\bra{-}\Big)
\end{equation}

\begin{figure}
\vspace{0mm}
\includegraphics[width=1 \columnwidth]{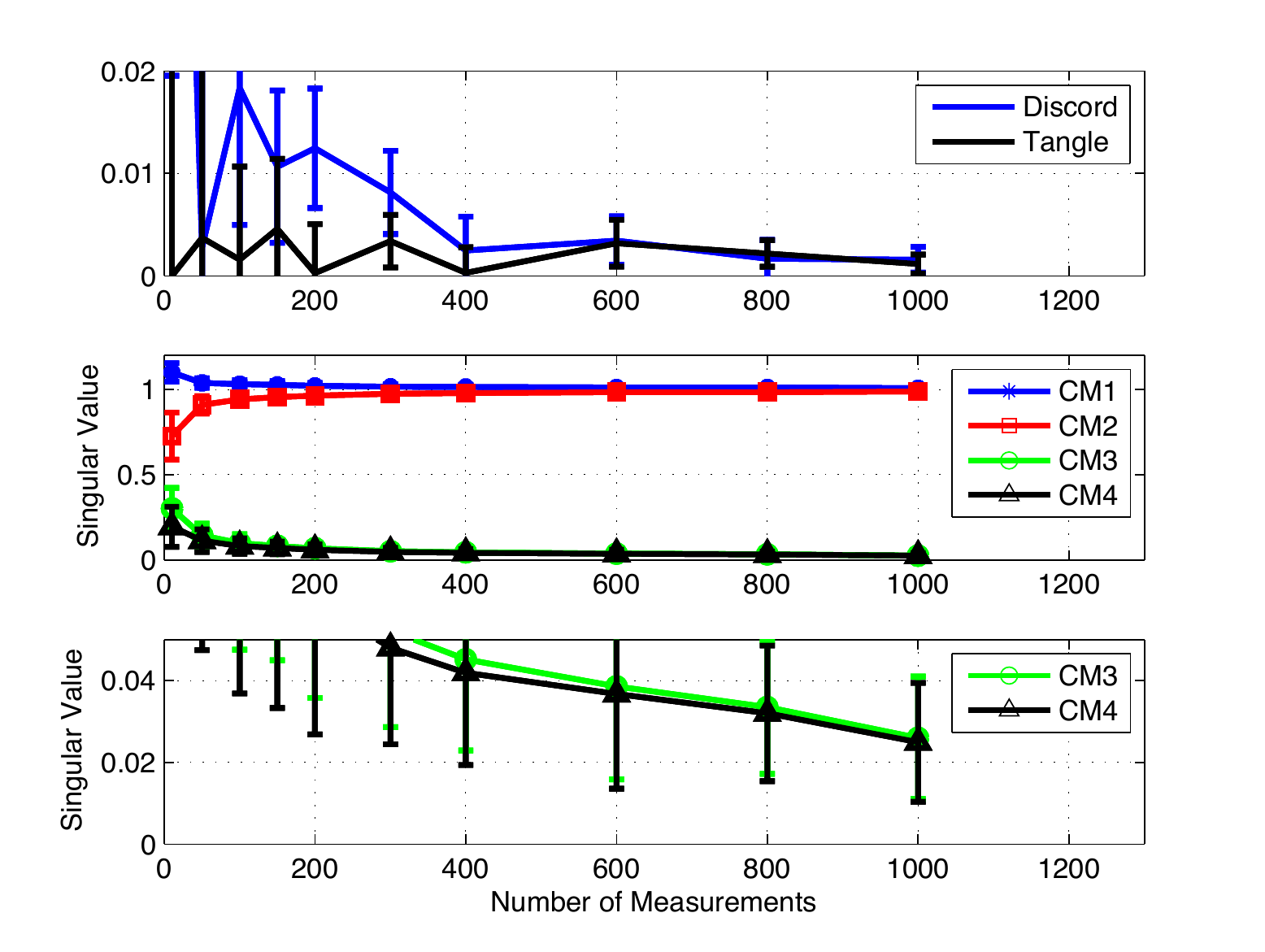}
\vspace{-4mm}
\caption{
Numerical simulation results showing the effect of taking a finite number of measurements on the properties of a tomographic reconstruction of $\rho_1$ (Eqn. (1)).  See text for more details. The following properties of the reconstructed density matrices are shown a) Discord and Tangle \cite{tangle}, which are ideally both zero. b) Correlation matrix singular values $[CM1,CM2,CM3,CM4]$, which are ideally $[1,1,0,0]$ c) Zoom in on $CM3$ and $CM4$ in b). Note that discord $D_{A}$ is presented and does not differ significantly from $D_{B}$.
}
\label{lines}
\vspace{-3mm}
\end{figure}

This state has zero discord when considering measurements on either qubit ($D_A{=}D_B{=}0$), zero tangle (a measure of two qubit entanglement \cite{tangle}) and correlation matrix singular values $[CM1,CM2,CM3,CM4]{=} [1,1,0,0]$ giving it a rank $R{=}2$. Figure \ref{lines} presents these properties of the noisy copies of the ideal state as a function of the number of simulated measurements. As one would expect, the more measurements taken the closer the properties of the reconstructed states are to those of the ideal state.

Fig.~\ref{lines}a shows that at least 1000 measurements are required before the discord and tangle are zero to within one standard deviation. In other words, if 800 measurements are taken per basis then -- even if a perfect separable state was made -- one would expect to reconstruct a state with non-zero entanglement beyond one standard deviation

Fig.~\ref{lines}b-c shows that the singular values of the correlation matrix are also sensitive, yielding a full rank ($R{=}4$) beyond one standard deviation even after 1000 measurements. One should therefore expect to obtain small but non-zero numbers for these values even when generating the ideal state in the lab. Note that at some point increasing the measurement number (and therefore experimental duration) can lead to other problems such as assuring that the identical state is prepared for longer times. 

\begin{figure}[h]
\vspace{5mm}
\includegraphics[width=1 \columnwidth]{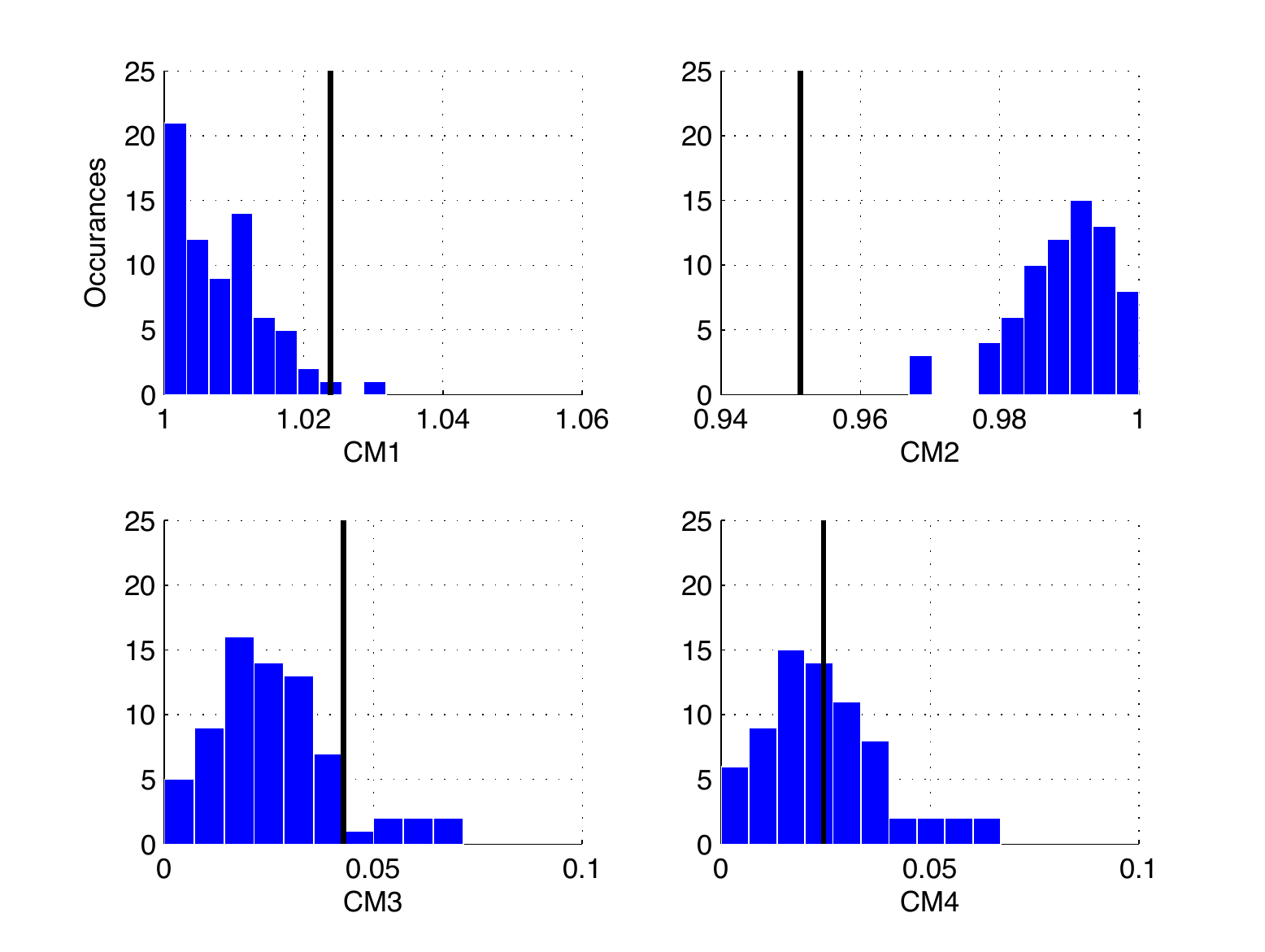}
\vspace{-4mm}
\caption{
Numerical simulation results showing distribution of correlation matrix singular values $[CM1,CM2,CM3,CM4]$ expected when performing maximum likelihood tomographic reconstruction of the state $\rho_1$ (Eqn.1) and using 1000 measurements per measurement basis. The distribution of correlation matrix singular values are shown as blue bars. Ideally the correlation matrix singular values are $[CM1{=}1,CM2{=}1,CM3{=}0,CM4{=}0]$. Solid black lines are the values calculated from the experimentally reconstructed density matrix for 1000 measurements. Note that the data shown in this figure was used to calculate the last point in Fig. \ref{lines}b).
}
\label{histograms}
\vspace{-3mm}
\end{figure}

Recall that each point in Fig.~\ref{lines} is the average outcome of 70 simulated experiments based on noisy copies of the ideal state. Figure \ref{histograms} shows the full distribution of each of the four correlation matrix singular values (CM1 - CM4) for 1000 simulated measurements of the ideal state. Also shown, as black lines, are the experimentally reconstructed maximum likelihood results for this state (also taken with 1000 measurements). The results show that the experimentally observed non-zero 3\textsuperscript{rd} and 4\textsuperscript{th} singular values (CM3 \& CM4) are entirely consistent with those expected for the ideal state. Differences in the 1\textsuperscript{st} and 2\textsuperscript{nd} singular values (CM1 \& CM2) can easily be attributed to a slight imbalancing of the, ideally equally weighted, two terms in $\rho_1$ (Eqn (1)). This could be caused by small errors in the pulse sequence used to generate this state.

\subsection{Separable operations can increase the correlation rank}
In this section we show that applying a separable operation to a bipartite density matrix can increase the rank of the correlation matrix \cite{DVB}, henceforth called the correlation rank. We consider a bipartite Hilbert space $\mathcal{H}=\mathcal{H}_A\otimes\mathcal{H}_B$ and sets of arbitrary quantum operations $\{\epsilon^A_i\}$, $\{\epsilon^B_i\}$ working on the subspaces $\mathcal{H}_A$ and $\mathcal{H}_B$, respectively. We call an operation $\epsilon_{\mathrm{sep}}$ on $\mathcal{H}$ \textit{separable} if it can be put into the form
$\epsilon_{\mathrm{sep}}=\sum_{i=1}^Pp_i\epsilon^A_i\otimes\epsilon^B_i$, with probabilities $p_i$ and $P$ denotes the number of terms in the sum. In order to introduce the notation which we will use in the following, we first review the definition of the correlation matrix. First, we represent the initial state $\rho$ in terms of arbitrary, fixed bases of Hermitian operators $\{A_i\}$ and $\{B_i\}$ on $\mathcal{H}_A$ and $\mathcal{H}_B$, respectively:
\begin{align}
\rho=\sum_{i=1}^{d_A^2}\sum_{j=1}^{d_B^2}r_{ij}A_i\otimes B_j,
\end{align}
where $d_A$ and $d_B$ denote the dimensions of $\mathcal{H}_A$ and $\mathcal{H}_B$. The correlation matrix $R=(r_{ij})$ can be decomposed into its singular value decomposition, using orthogonal matrices $U$ and $V$: $R=U\mathrm{diag}(c_1,\dots,c_L,0,\dots)V^T$. Here, $c_i$ represent the nonzero singular values and $L$ defines the correlation rank. Introducing $S_i=\sum_ju_{ji}A_j$ and $F_i=\sum_iv_{ji}B_j$, where $U=(u_{ij})$ and $V=(v_{ij})$, we can rewrite $\rho$ as \cite{DVB}
\begin{align}
\rho=\sum_{i=1}^Lc_iS_i\otimes F_i.
\end{align}

In Ref.~\cite{TURKU} it was shown that  \textit{unilocal} operations, defined as operations of the form $\epsilon^A\otimes\mathbb{I}$, cannot increase the correlation rank. First of all, we assure that this also holds for a \textit{bilocal} operation $\epsilon^A\otimes\epsilon^B$, which requires access to both subsystems.  We find
\begin{align}
\rho'=(\epsilon^A\otimes\epsilon^B)\rho=\sum_{i=1}^L\sum_{k=1}^{d_A^2}\sum_{j=1}^{d_B^2}c_id_{ik}e_{ij}A_k\otimes B_j,
\end{align}
with $\epsilon_A(S_i)=\sum_kd_{ik}B_k$ and $\epsilon_B(F_i)=\sum_je_{ij}B_j$. The rank of $\rho'$ is now given by the rank of the matrix $F=(f_{kj})$, with $f_{kj}=\sum_{i=1}^Lc_id_{ik}e_{ij}$ which is still limited by the minimum of the rank of the three matrices $C=\text{diag}(c_1,\dots,c_L,0,\dots)$, $D=(d_{ik})$, and $E=(e_{ij})$, hence, by $L$. Hence, even bilocal operations cannot increase the correlation rank. However, if we allow for arbitrary separable maps the correlation rank can be increased:
\begin{align}
\rho'=\sum_{i=1}^Pp_i(\epsilon_i^A\otimes\epsilon_i^B)\rho=\sum_{i=1}^P\sum_{j=1}^L\sum_{k=1}^{d_A^2}\sum_{l=1}^{d_B^2}p_ic_jd^i_{jk}e^i_{jl}A_k\otimes B_l.
\end{align}
We have introduced $\epsilon_i^A(S_j)=\sum_{k=1}^{d_A^2}d^i_{jk}A_k$ and $\epsilon_i^B(F_j)=\sum_{l=1}^{d_B^2}e^i_{jl}B_l$. Now, the rank of the correlation matrix $f_{kl}=\sum_{i=1}^P\sum_{j=1}^Lp_ic_jd^i_{jk}e^i_{jl}$ can in principle go up to $\text{min}\{P\cdot L,d_A^2,d_B^2\}$.

\subsection{Correlated dephasing on two qubits}
We assume a state of two qubits with reduced Bloch vectors $\vec{r}^A$ and $\vec{r}^B$, expressing the reduced density matrices of subsystems $\mathcal{H}_A$ and $\mathcal{H}_B$ as
\begin{align}
\rho_{A,B}=\frac{1}{2}\left(\mathbb{I}+\vec{r}^{A,B}\cdot\vec{\sigma}\right).
\end{align}
Here $\vec{\sigma}$ denotes the vector of Pauli-matrices $\vec{\sigma}=(\sigma_x,\sigma_y,\sigma_z)^T$. The correlations between the qubits is contained in a real-valued matrix $\beta=(\beta_{ij})$, such that the total state can be represented in terms of Pauli matrices and the identity in its Fano-Form (see, e.g., \cite{Geometry}):
\begin{align}\label{eq.fanoform}
\rho=\frac{1}{4}\left(\mathbb{I}\otimes\mathbb{I}+\sum_{i=1}^{3}r^A_i\sigma_i\otimes\mathbb{I}+\sum_{i=1}^{3}r^B_i\mathbb{I}\otimes\sigma_i+\sum_{i=1}^{3}\sum_{j=1}^{3}\beta_{ij}\sigma_i\otimes\sigma_j\right).
\end{align}
The correlation matrix is given by
\begin{align}\label{eq.corrmatrix}
R=\frac{1}{4}\begin{pmatrix}
1 & \left(\vec{r}^B\right)^T\\
\vec{r}^A & \beta
\end{pmatrix},
\end{align}
where $\vec{r}^T$ denotes the $1\times 3$ matrix $\vec{r}^T=(r_x,r_y,r_z)$. The rank $L$ of the correlation matrix $R$ is given by \cite{MeyerJr}
\begin{align}\label{eq.rankformula}
L=1+\text{rk}(\beta-\vec{r}^A\otimes\vec{r}^B).
\end{align}
Here $\vec{r}^A\otimes\vec{r}^B$ denotes the outer product: $(\vec{r}^A\otimes\vec{r}^B)_{ij}=r^A_ir^B_j$.
In general, the matrix $\beta$ can be decomposed into singular values, $\beta_{ij}=\sum_{k=1}^3v_{ik}d_kw_{jk}$, with orthogonal matrices $V=(v_{ij})$ and $W=(w_{ij})$. Introducing two sets of orthonormal vectors $\left(\vec{v}_k\right)_i=v_{ik}$ and $\left(\vec{w}_k\right)_j=w_{jk}$ as the columns of $V$ and $W$, we rewrite the state $\rho$ as
\begin{align}\label{eq.fano}
\rho=\frac{1}{4}\left(\mathbb{I}\otimes\mathbb{I}+\vec{r}^A\cdot\vec{\sigma}\otimes\mathbb{I}+\mathbb{I}\otimes\vec{r}^B\cdot\vec{\sigma}+\sum_{k=1}^{3}d_k\vec{v}_k\cdot\vec{\sigma}\otimes\vec{w}_k\cdot\vec{\sigma}\right).
\end{align}

\subsubsection{Dephasing by fluctuating rotations}
Single-qubit rotations of a Bloch vector $\vec{r}$ around an axis determined by the normalized vector $\vec{n}$ are described by the operator $R_{\vec{n}}(\theta)=e^{-i\theta\vec{n}\cdot\vec{\sigma}/2}$. Using the relation
\begin{align}
(\vec{n}\cdot\vec{\sigma})(\vec{r}\cdot\vec{\sigma})=(\vec{n}\cdot\vec{r})\mathbb{I}+i(\vec{n}\times\vec{r})\cdot\vec{\sigma},
\end{align}
we obtain
\begin{align}
(\vec{n}\cdot\vec{\sigma})(\vec{r}\cdot\vec{\sigma})(\vec{n}\cdot\vec{\sigma})&=(\vec{n}\cdot\vec{r})(\vec{n}\cdot\vec{\sigma})-(\vec{n}\times\vec{r}\times\vec{n})\cdot\vec{\sigma}\notag\\
&=2(\vec{n}\cdot\vec{r})(\vec{n}\cdot\vec{\sigma})-(\vec{r}\cdot\vec{\sigma}).
\end{align}
In combination with $e^{-i\theta\vec{n}\cdot\vec{\sigma}/2}=\cos(\theta/2)\mathbb{I}+i\sin(\theta/2)\vec{n}\cdot\vec{\sigma}$, this allows us to describe an arbitrary rotation of a Bloch vector $\vec{r}$ around $\vec{n}$ by
\begin{widetext}
\begin{align}
e^{-i\theta\vec{n}\cdot\vec{\sigma}/2}\vec{r}\cdot\vec{\sigma}e^{i\theta\vec{n}\cdot\vec{\sigma}/2}&=\left[\cos^2\left(\frac{\theta}{2}\right)-\sin^2\left(\frac{\theta}{2}\right)\right]\vec{r}\cdot\vec{\sigma}-2i\sin\left(\frac{\theta}{2}\right)\cos\left(\frac{\theta}{2}\right)(\vec{r}\times\vec{n})\cdot\vec{\sigma}\notag\\&\quad+2\sin^2\left(\frac{\theta}{2}\right)(\vec{n}\cdot\vec{r})\vec{n}\cdot\vec{\sigma}.
\end{align}
\end{widetext}
Averaging uniformly over the angle $\theta$ generates a dephasing effect. The new Bloch vector points into the direction $\vec{n}$ of the rotation:
\begin{align}
\frac{1}{2\pi}\int_{0}^{2\pi}d\theta e^{-i\theta\vec{n}\cdot\vec{\sigma}/2}\vec{r}\cdot\vec{\sigma}e^{i\theta\vec{n}\cdot\vec{\sigma}/2}=(\vec{n}\cdot\vec{r})\vec{n}\cdot\vec{\sigma}.
\end{align}

\subsubsection{Correlated dephasing}
A separable map describing correlated dephasing in both subsystems is generated by
\begin{widetext}
\begin{align}
\epsilon^{\vec{n}}_{cd}(\vec{v}\cdot\vec{\sigma}\otimes\vec{w}\cdot\vec{\sigma})&=\frac{1}{2\pi}\int_{0}^{2\pi}d\theta R_{\vec{n}}(\theta)\vec{v}\cdot\vec{\sigma}R^{\dagger}_{\vec{n}}(\theta)\otimes R_{\vec{n}}(\theta)\vec{w}\cdot\vec{\sigma}R^{\dagger}_{\vec{n}}(\theta)\notag\\
&=\frac{1}{2\pi}\int_{0}^{2\pi}d\theta e^{-i\theta\vec{n}\cdot\vec{\sigma}/2}\vec{v}\cdot\vec{\sigma}e^{i\theta\vec{n}\cdot\vec{\sigma}/2}\otimes e^{-i\theta\vec{n}\cdot\vec{\sigma}/2}\vec{w}\cdot\vec{\sigma}e^{i\theta\vec{n}\cdot\vec{\sigma}/2}\notag\\
&=\frac{1}{2}\vec{v}\cdot\vec{\sigma}\otimes\vec{w}\cdot\vec{\sigma}-\frac{1}{2}\vec{v}\cdot\vec{\sigma}\otimes(\vec{n}\cdot\vec{w})\vec{n}\cdot\vec{\sigma}-\frac{1}{2}(\vec{n}\cdot\vec{v})\vec{n}\cdot\vec{\sigma}\otimes\vec{w}\cdot\vec{\sigma}\notag\\
&\quad-\frac{1}{2}(\vec{v}\times\vec{n})\cdot\vec{\sigma}\otimes(\vec{w}\times\vec{n})\cdot\vec{\sigma}+\frac{3}{2}(\vec{n}\cdot\vec{v})\vec{n}\cdot\vec{\sigma}\otimes(\vec{n}\cdot\vec{w})\vec{n}\cdot\vec{\sigma}.
\end{align}
\end{widetext}

\subsubsection{Kraus representation}
Correlated dephasing can be given in form of a Kraus representation:
\begin{align}
\epsilon^{\vec{n}}_{cd}(\rho)=\frac{1}{2}K_1\rho K_1+\frac{1}{4}K_2\rho K_2+\frac{1}{4}K_3\rho K_3,
\end{align}
with the self-adjoint Kraus operators
\begin{align}
K_1&=\frac{1}{\sqrt{2}}\left(-\mathbb{I}\otimes\mathbb{I}+\vec{n}\cdot\vec{\sigma}\otimes\vec{n}\cdot\vec{\sigma}\right)\notag\\
K_2&=\frac{1}{\sqrt{2}}\left(\mathbb{I}\otimes\mathbb{I}+\vec{n}\cdot\vec{\sigma}\otimes\vec{n}\cdot\vec{\sigma}\right)\notag\\
K_3&=\frac{1}{\sqrt{2}}\left(-\mathbb{I}\otimes\vec{n}\cdot\vec{\sigma}+\vec{n}\cdot\vec{\sigma}\otimes\mathbb{I}\right).
\end{align}

\subsubsection{Initial rank-1 states}

We show that $L=1$ if and only if $\rho=\rho_A\otimes\rho_B$: First, if $\rho$ is a product state then obviously $L=1$, as $\rho_A$ and $\rho_B$ can be seen as elements of operator bases of $\mathcal{H}_A$ and $\mathcal{H}_B$, respectively. Conversely, assume that $L=1$, then, according to Eq.~(\ref{eq.corrmatrix}), we find that $(1,\vec{r}^B)=c_1(r^A_1,\vec{\beta}_1)=c_2(r^A_2,\vec{\beta}_2)=c_3(r^A_3,\vec{\beta}_3)\in\mathbb{R}^4$ with $\beta=(\vec{\beta}_1,\vec{\beta}_2,\vec{\beta}_3)$. From this we conclude that $c_i=1/r^A_i$ and $\vec{\beta}_i=r^A_i\vec{r}^B$. Inserting this into Eq.~(\ref{eq.fanoform}) yields
\begin{align}
\rho&=\frac{1}{4}\left(\mathbb{I}\otimes\mathbb{I}+\vec{r}^A\cdot\vec{\sigma}\otimes\mathbb{I}+\mathbb{I}\otimes\vec{r}^B\cdot\vec{\sigma}+\vec{r}^A\cdot\vec{\sigma}\otimes\vec{r}^B\cdot\vec{\sigma}\right)\notag\\
&=\frac{1}{2}(\mathbb{I}+\vec{r}^A\cdot\vec{\sigma})\otimes\frac{1}{2}(\mathbb{I}+\vec{r}^B\cdot\vec{\sigma}).
\end{align}
This concludes the proof and implies that $\text{rk}(\beta)\leq1$ for product states. Moreover, if $\text{rk}(\beta)=1$ then the left- and right-singular vectors of $\beta$ are given by the reduced Bloch vectors $\vec{r}^A$ and $\vec{r}^B$, respectively and $\beta=\vec{r}^A\otimes\vec{r}^B$, in agreement with Eq.~(\ref{eq.rankformula}).

We now determine the rank of the product state after being subject to correlated dephasing along $\vec{n}$. The correlation matrix of the state after application of the map is given by
\begin{align}
\epsilon^{\vec{n}}_{cd}(R)=\frac{1}{4}\left(
\begin{array}{c|c}
1 & (\vec{r}^B\cdot\vec{n})\vec{n}^T \\\hline
(\vec{r}^A\cdot\vec{n})\vec{n} &\epsilon^{\vec{n}}_{cd}(\vec{r}^A\otimes\vec{r}^B)
\end{array}\right),
\end{align}
with the rank
\begin{widetext}
\begin{align}
\text{rk}(\epsilon^{\vec{n}}_{cd}(R))&=1+\text{rk}\left(\epsilon^{\vec{n}}_{cd}(\vec{r}^A\otimes\vec{r}^B)-(\vec{r}^B\cdot\vec{n})(\vec{r}^B\cdot\vec{n})\vec{n}\otimes\vec{n}\right)\notag\\
&=1+\text{rk}\left(\frac{1}{2}\left(\vec{r}^A\otimes\vec{r}^B-\vec{r}^A\otimes(\vec{r}^B\cdot\vec{n})\vec{n}-(\vec{r}^A\cdot\vec{n})\vec{n}\otimes\vec{r}^B-(\vec{r}^A\times\vec{n})\otimes(\vec{r}^B\times\vec{n})\right.\right.\notag\\&\left.\left.\quad+(\vec{r}^B\cdot\vec{n})(\vec{r}^B\cdot\vec{n})\vec{n}\otimes\vec{n}\right)\right)\notag\\
&=1+\text{rk}\left(\frac{1}{2}\left((\vec{r}^A-(\vec{r}^A\cdot\vec{n})\vec{n})\otimes(\vec{r}^B-(\vec{r}^B\cdot\vec{n})\vec{n})-(\vec{r}^A\times\vec{n})\otimes(\vec{r}^B\times\vec{n})\right)\right).
\end{align}
\end{widetext}
We obtain $\text{rk}(\epsilon^{\vec{n}}_{cd}(R))=1$ if and only if $\vec{n}=\vec{r}^A/r^A$ or $\vec{n}=\vec{r}^B/r^B$, with $r^A=\sqrt{\vec{r}^A\cdot\vec{r}^A}$ and $r^B=\sqrt{\vec{r}^B\cdot\vec{r}^B}$. Otherwise, the final rank is 3.

\subsubsection{Initial rank-2 states}
We consider states with maximally mixed reduced density matrices, i.e., $\vec{r}^A=\vec{r}^B=\vec{0}$. The rank of the correlation matrix is then given by $L=\text{rk}(\beta)+1$. Hence, classical states with maximally mixed marginals have $\text{rk}(\beta)\leq1$ and $\beta=\textbf{0}$ corresponds to the overall maximally mixed state, which has $L=1$. Let's assume we have $L=2$, hence, only one singular value of $\beta$ is nonzero. Such a state has always zero discord as it can be written as
\begin{align}\label{eq.ccmmixedreducedstates}
\rho_0=\frac{1}{4}\left(\mathbb{I}\otimes\mathbb{I}+d\vec{v}\cdot\vec{\sigma}\otimes\vec{w}\cdot\vec{\sigma}\right).
\end{align}
The matrix $\beta$ is given by the outer product $\beta=d\vec{v}\otimes\vec{w}$.

We will use the following relation:
\begin{align}\label{eq.rankrelation}
\text{rk}\left(\sum_{k=1}^3\vec{v}_k\otimes\vec{w}_k\right)=\text{rk}(VW^T)\leq\text{min}\{\text{rk}V,\text{rk}W\},
\end{align}
where the rank of the matrices $V$ and $W$ is determined by the number of linear independent vectors $\vec{v}_k$ and $\vec{w}_k$, respectively. If one of the two sets is linearly independent, the rank is given by the number of linear independent vectors of the other set. Using this it is possible to specify a combination of conditions which guarantee the conversion of a classical rank-2 state with maximally mixed marginals into a fully correlated rank-4 state by correlated dephasing. Correlated dephasing on (\ref{eq.ccmmixedreducedstates}) yields the state
\begin{widetext}
\begin{align}
\epsilon^{\vec{n}}_{cd}(\rho_0)&=\frac{1}{4}\left(\mathbb{I}\otimes\mathbb{I}+\frac{d}{2}\left[\vec{v}\cdot\vec{\sigma}\otimes(\vec{w}-(\vec{n}\cdot\vec{w})\vec{n})\cdot\vec{\sigma}+(\vec{n}\cdot\vec{v})\vec{n}\cdot\vec{\sigma}\otimes(3(\vec{n}\cdot\vec{w})\vec{n}-\vec{w})\cdot\vec{\sigma}\right.\right.\notag\\
&\hspace{2.3cm}\left.\left.\quad-(\vec{v}\times\vec{n})\cdot\vec{\sigma}\otimes(\vec{w}\times\vec{n})\cdot\vec{\sigma}\right]\right).
\end{align}
\end{widetext}
The set $\{\vec{v},(\vec{n}\cdot\vec{v})\vec{n},-\vec{v}\times\vec{n}\}$ is linearly independent if and only if $\vec{n}$ is neither equal nor orthogonal to $\vec{v}$ (or put equivalently $0<\vec{n}\cdot\vec{v}<1$, since both $\vec{v}$ and $\vec{n}$ are normalized to one). The latter implies that $\vec{n}$ should be different from all the left-singular vectors of $\beta$, even those with singular value zero. According to Eq.~(\ref{eq.rankrelation}), the rank of $\beta$ after the correlated dephasing is now given by the number of linearly independent vectors in the set $\{\vec{w}-(\vec{n}\cdot\vec{w})\vec{n},3(\vec{n}\cdot\vec{w})\vec{n}-\vec{w},\vec{w}\times\vec{n}\}$. Again, all three vectors are linearly independent if and only if $0<\vec{n}\cdot\vec{w}<1$. A rank-2 state with maximally mixed marginals is therefore converted into a rank-4 state if and only if $0<\vec{n}\cdot\vec{v}<1$ and $0<\vec{n}\cdot\vec{w}<1$. For states which are symmetric under permutation of the two qubits ($\vec{v}=\vec{w}$), this can be expressed directly in terms of $\beta$ as the condition
\begin{align}
0<\vec{n}^T\beta\vec{n}/d<1,
\end{align}
where the Hilbert Schmidt norm of $\beta$ is given by $d=\|\beta\|=\sqrt{\text{Tr}\beta^T\beta}$.

The above analysis tells us how to create rank-4 states. When do we obtain a state of rank 1, 2 or 3? If $\vec{n}=\vec{v}$ we get
\begin{align}
\epsilon^{\vec{n}}_{cd}(\rho_0)&=\frac{1}{4}\left(\mathbb{I}\otimes\mathbb{I}-d(\vec{v}\cdot\vec{w})\vec{v}\cdot\vec{\sigma}\otimes\vec{v}\cdot\vec{\sigma}\right),
\end{align}
which has rank 2 if $\vec{v}\cdot\vec{w}>0$, otherwise rank 1. The same holds for $\vec{n}=\vec{w}$. For $\vec{n}\cdot\vec{v}=0$ the final state
\begin{widetext}
\begin{align}
\epsilon^{\vec{n}}_{cd}(\rho_0)&=\frac{1}{4}\left(\mathbb{I}\otimes\mathbb{I}+\frac{d}{2}\left[\vec{v}\cdot\vec{\sigma}\otimes(\vec{w}-(\vec{n}\cdot\vec{w})\vec{n})\cdot\vec{\sigma}-(\vec{v}\times\vec{n})\cdot\vec{\sigma}\otimes(\vec{w}\times\vec{n})\cdot\vec{\sigma}\right]\right)
\end{align}
\end{widetext}
has rank 3 as long as $\vec{n}\cdot\vec{w}<1$ (equivalent to $\vec{n}\neq\vec{w}$), which is also the case for $\vec{n}\cdot\vec{w}=0$ if $\vec{n}\cdot\vec{v}<1$.

\subsection{Summary}
For initial rank-1 states $\rho=\rho_A\otimes\rho_B$, with nonzero reduced Bloch vectors $\vec{r}^A$ and $\vec{r}^B$, correlated dephasing around $\vec{n}$ yields a state of
\begin{enumerate}
\item[L=1] if $\vec{n}$ is equal to either $\vec{r}^A$ or $\vec{r}^B$,
\item[L=3] if $\vec{n}$ differs from both $\vec{r}^A$ and $\vec{r}^B$.
\end{enumerate}

For initial rank-2 states with maximally mixed reduced density matrices $(\vec{r}^A=\vec{r}^B=0)$, correlated dephasing around $\vec{n}$ yields a state of
\begin{enumerate}
\item[L=1] if $\vec{n}=\vec{v}$ ($\vec{n}=\vec{w}$) and $\vec{n}\perp\vec{w}$ ($\vec{n}\perp\vec{v}$),
\item[L=2] if $\vec{n}=\vec{v}$ ($\vec{n}=\vec{w}$) and $\vec{n}\not\perp\vec{w}$ ($\vec{n}\not\perp\vec{v}$),
\item[L=3] if $\vec{n}\neq\vec{w}$ ($\vec{n}\neq\vec{v}$) and $\vec{n}\perp\vec{v}$ ($\vec{n}\perp\vec{w}$),
\item[L=4] if $\vec{n}$ is neither equal nor orthogonal to and $\vec{v}$ and $\vec{w}$,
\end{enumerate}
where $\vec{v}$ and $\vec{w}$ denote the left- and right-singular vectors of the matrix $\beta=d\vec{v}\otimes\vec{w}$ in Eq.~(\ref{eq.fanoform}). These results are summarized in Tab.~\ref{tab.rank}. For the special case of $\vec{v}=\vec{w}$, we obtain $L=4$ if and only if
\begin{align}
0<\frac{\vec{n}^T\beta\vec{n}}{\sqrt{\text{Tr}\beta^T\beta}}<1.
\end{align}

The present analysis can be extended beyond qubit-systems based on the definitions of the Fano form, rotations and generalized Bloch vectors for higher dimensions \cite{Geometry}.

\begin{figure}
\begin{tabular}{l|ccc}
 & $\vec{n}=\vec{w}$ & $\quad\vec{n}\cdot\vec{w}=0\quad$ & $0<\vec{n}\cdot\vec{w}<1$\\\hline
 $\vec{n}=\vec{v} $& 2 & 1 & 2\\
 $\vec{n}\cdot\vec{v}=0$ & 1 & 3 & 3\\
 $0<\vec{n}\cdot\vec{v}<1$ & 2 & 3 & 4
\end{tabular}
\caption{Rank of the correlation matrix after application of correlated dephasing to a rank-2 state with maximally mixed marginals.}
\label{tab.rank}
\end{figure}

\end{document}